\documentclass[twocolumn,aps,prl,showpacs]{revtex4}

\usepackage{graphicx}
\usepackage{graphics}
\usepackage{epsfig}

\begin{document}
\title{{\bf\Large Anomalous decay of pion and eta at finite temperature}}
\author{\bf P. Costa}
\email{pcosta@teor.fis.uc.pt}
\author{\bf M. C. Ruivo}
\email{maria@teor.fis.uc.pt}
\affiliation{Departamento de F\'{\i}sica, Universidade de Coimbra, P-3004-516 Coimbra, Portugal} 
\author{\bf Yu. L. Kalinovsky\footnote{Permanent Address: Laboratory of Information Technologies, Joint Institute for Nuclear Research, Dubna, Russia}}
\email{kalinov@qcd.phys.ulg.ac.be}
\affiliation{Universit\'{e} de Li\`{e}ge, D\'{e}partment de Physique B5, Sart Tilman, B-4000, LIEGE 1, Belgium}
\date{\today}
\begin{abstract}
We study the anomalous decays $\pi^0,\, \eta \rightarrow\gamma\gamma$ in the framework of the three--flavor Nambu--Jona-Lasinio [NJL] model at finite temperature.  The similarities and differences between these results and those obtained at zero temperature and finite density are discussed. In both cases the lifetimes of these mesons decrease significantly at the critical point, although this might not be sufficient to   observe enhancement of these decays in heavy-ion collisions.
\end{abstract}
%
\maketitle


\section{Introduction}

The investigation of  pseudoscalar meson observables at high density or temperature is specially relevant since, due the fact that the origin of these mesons is associated to the dynamical or explicit breaking of symmetries,  possible   modifications in the mass spectrum, lifetime and widths  of those mesons could provide signatures for the restoration of symmetries  associated to phase transitions that are expected to take place under such extreme conditions \cite{kanaya,rhic}.  

A great deal of attention has been dedicated to the electromagnetic decays of the neutral pseudoscalar  mesons $\pi^0$ and $\eta$, and  calculations of  such observables in the framework of different models  may be found in the literature \cite{cleo,tdavid,tklev,tpisarski,dorokhov,oka,hashimoto,costaD,caldas}.  Since the great percentage of photons in the background of heavy-ion collisions is due to the decays $\pi^0 (\eta)\,\rightarrow \gamma\gamma$, these  processes  deserve   special attention. On the other side, modifications of anomalous mesonic interactions  is a topic that has attracted lot of interest, since,  while for fermions the axial anomaly is not affected by the medium, the opposite is expected for anomalous mesonic interactions \cite{tpisarski}. 
Whether enhancement or suppression of $\pi^0 (\eta)\,\rightarrow \gamma\gamma$  will be found in experiments  of heavy-ion collisions  is an open question. Although the lifetimes of these neutral mesons are much longer than hadronic time scales and its decays  might  not be observed inside the fireball,  it is nevertheless important to understand the underlying physics, that   is probably the same of other more complex anomalous mesonic decays ($\omega\rightarrow \pi\pi\pi\,, \omega \rightarrow \rho\pi$) that are relevant for experiments in the hot/dense region \cite{tpisarski}.

Temperature effects on the process $\pi^0 \rightarrow \gamma\gamma$ have been object of several studies but less attention has been given to the investigation of  $\eta \rightarrow \gamma\gamma$.  Calculations  of the $\pi^0\rightarrow \gamma\gamma$ \cite{hashimoto,tklev,tdavid} in the framework of the $SU(2)$ Nambu-Jona-Lasinio [NJL] \cite{NJL} model indicate an enhancement of the width around the temperature at which the pion dissociates into $\bar q q$ pairs (Mott temperature).
 
The calculation of temperature effects on the   decays $\pi^0 (\eta) \rightarrow \gamma\gamma$  is important in several ways,  but the calculation for the $\eta$ is more involved, since the structure of this meson exhibits a mixing of strange and non strange quarks. The welknown fact that  chiral symmetry shows a tendency to be restored in the non strange sector, while the same is not evident for the strange sector, makes the study of the electromagnetic $\eta$ decay  particularly interesting. The present work comes in the sequel of our previous studies on the behavior of neutral mesons in hot and dense matter \cite{costaI,costaB,costabig,costaD}, in the framework of the $SU(3)$ NJL model \cite{njlt,RuivoSousa}. 
In Ref. \cite{costaD} we have studied the decays $\pi^0 (\eta)\,\rightarrow \gamma\gamma$, in vacuum and in quark matter in $\beta$ --equilibrium at zero temperature and we  have shown that these decays are  affected by the medium, their lifetime decreasing significantly at the critical density.  It should be noticed that, although the behavior of several observables with temperature is qualitatively similar to that with density, there are fundamental differences between   media with $T\neq 0\,, \rho=0$ and $T= 0\,, \rho\neq 0$. In the first case the phase transition is  probably a smooth crossover and there are threshold effects on the mesons, that dissociate in $\bar q q$ pairs at the Mott temperature; in the second case the phase transition is first order and the threshold effects mentioned above do not generally take place. The question whether these differences will affect significantly the decays $\pi^0 (\eta) \rightarrow \gamma\gamma$ should be clarified.
 
The aim of this work is to report an investigation on the anomalous decays of $\pi^0$ and $\eta$ at finite temperature and zero density, in comparison with the behavior at finite density and zero temperature, giving special attention to the possible enhancement or suppression of these decays at the respective Mott temperatures.  


\section{$\pi^0(\eta)\rightarrow \gamma\gamma$ decays at finite temperature}
        
We perform our calculations in the framework of the  three--flavor NJL model, including the determinantal 't Hooft interaction that breaks the $U_A(1)$ symmetry, that  has the  following Lagrangian: 
\begin{eqnarray}
{\mathcal L\,}&=& \bar q\,(\,i\, {\gamma}^{\mu}\,\partial_\mu\,-\,\hat m)\,q\nonumber\\
&+& \frac{1}{2}\,g_S\,\,\sum_{a=0}^8\, [\,{(\,\bar q\,\lambda^a\, q\,)}
^2\,\,+\,\,{(\,\bar q \,i\,\gamma_5\,\lambda^a\, q\,)}^2\,] \nonumber\\
&+& g_D\,\{\mbox{det}\,[\bar q\,(1+\gamma_5)\,q] +\mbox{det}
\,[\bar q\,(1-\gamma_5)\,q]\}. \label{1} 
\end{eqnarray}
Here $q = (u,d,s)$ is the quark field with three flavors, $N_f=3$, and three colors, $N_c=3$. $\hat{m}=\mbox{diag}(m_u,m_d,m_s)$ is the current quark mass matrix and $\lambda^a$ are the Gell--Mann matrices, a = $0,1,\ldots , 8$, ${ \lambda^0=\sqrt{\frac{2}{3}} \, {\bf I}}$.

By using  a standard bosonization procedure,  an effective meson action is obtained. The first variation of the action leads to the  the gap equations, which gives us the constituent quark masses, and the  second order expansion of the action over the meson fields  allows to obtain the meson propagators, from which several observables are calculated (for details see Refs. \cite{costaI,costaB,costabig,costaD}).

Our model parameters, the bare quark masses $m_d=m_u, m_s$, the coupling constants and the cutoff in three--momentum space, $\Lambda$, are  fitted to the experimental values of masses for pseudoscalar mesons ($M_{\pi^0} = 135.0$ MeV, $M_{K} = 497.7$ MeV) and $f_\pi = 92.4$ MeV, using  the parameterization of Refs. \cite{RKS,costaB,costabig}.

For the calculation of the   decays $H \longrightarrow \gamma 
\gamma$ we consider the appropriate triangle diagrams   \cite{costaD}. 
The corresponding invariant amplitudes are given by: 
\begin{eqnarray}
&&{\tilde{\mathcal T}}_{H\rightarrow\gamma\gamma}(P,q_1,q_2) =\nonumber \\
&& i  \int \frac{d^4 p}{(2\pi)^4}\textrm{Tr} \left\{ \Gamma_H S(p - q_1) 
\hat{\epsilon}_1 S(p)  \hat{\epsilon}_2 S(p + q_2) \right\} \nonumber \\ &&
+ \textrm{exchange}. \label{trian}
\end{eqnarray}
Here the trace $\textrm{Tr}=\textrm{tr}_c\textrm{tr}_f\textrm{tr}_\gamma$, must be performed over color, flavor  and spinor indices. The meson vertex function, $\Gamma_H $,  has the $i \gamma_5$ form in the Dirac space, contains the corresponding coupling constant $g_{H\bar{q}q}$ (see Ref. \cite{costaD} for details) and is a  $3 \times 3$ matrix in the flavor space. $S(p)$ is the quark propagator $S(p)=\mbox{diag}(S_u, S_d, S_s)$, 
$\hat{\epsilon}_{1,2}$ is the photon polarization vector with momentum $q_{1,2}$.
The trace over flavors leads to different factors for different mesons,   $Q_{H\bar{q}q}$, that depend  on the electric charges and flavor of quarks into each meson $H$: 
$Q_\pi=1/3$, $Q_{\eta_u}=5/9$ and $Q_{\eta_s}=-\sqrt{2}/9$. 

We perform the calculation in the meson rest frame and use the kinematics $P=q_1+q_2$ and $P=(M_{H},\textbf{0})$.
Taking the trace in (\ref{trian}) we  obtain:
\begin{eqnarray}
    {\tilde{\mathcal T}}_{H\rightarrow\gamma\gamma}(P,q_1,q_2) = 	\epsilon_{\mu\nu\alpha\beta}\epsilon^\mu_1\epsilon^\nu_2 q_1^\alpha q_2^\beta  \, {\mathcal T}_{H\rightarrow\gamma\gamma} (P^2,q_1^2,q_2^2), 
\end{eqnarray}
where 
\begin{eqnarray}
    {\mathcal T}_{\pi^0\rightarrow\gamma\gamma} (P^2=M_{\pi^0}^2,q_1^2,q_2^2) =
    32\alpha\pi g_{\pi^0\bar{u}u}M_u 
    I^u_{\pi^0} 
\end{eqnarray}
and
\begin{eqnarray}
	& &{\mathcal T}_{\eta\rightarrow\gamma\gamma} (P^2=M_{\eta}^2,q_1^2,q_2^2) = \nonumber \\
	& &\frac{32\alpha\pi}{3\sqrt{3}}\bigl[\mbox{cos} \theta  (5 g_{\eta\bar{u}u}M_uI_\eta^u - 2 			 g_{\eta\bar{s}s}M_sI_\eta^s )\nonumber \\
	& &	- \mbox{sin} \theta  \sqrt{2} (5 g_{\eta\bar{u}u}M_uI_\eta^u +
		 g_{\eta\bar{s}s}M_sI_\eta^s )
	\bigr]. 
\end{eqnarray}
$\alpha$ is the fine structure constant and the integrals $I_{H}^i\equiv I_{H}^i(P) $ are  given by
\begin{eqnarray} \label{inth}
&&I_{H}^i(P)= \nonumber \\
&&i\int\frac{d^{4}p}{(2\pi)^{4}}\frac{1}{(p^{2}-M_{i}^{2})%
[(p-q_1)^{2}-M_{i}^{2}][(p+q_2)^{2}-M_{i}^{2}]},\nonumber \\
\end{eqnarray}
with $i=u,s$. At finite temperature, $I_H^i(P)$ takes the form: 
\begin{eqnarray}
I_{H}^i(P_0,\textbf{P}=0)&=& 
-\frac{1}{4\pi^2}\int_{0}^{\infty}dp\frac{p}{E_{i}^2}\frac{1}{4E_{i}^2-P_0^2}
\nonumber \\
&&\times\textrm{ln}\left(\frac{E_{i}+p}{M_i}\right)\left[1-2f(E_i(p)) \right],  \label{i3}
\end{eqnarray}
where $E_i$ is the quark energy.

Finally, the decay width is of the form:
\begin{equation}
\Gamma_{H\rightarrow\gamma\gamma}=\frac{M_{H}^{3}}{64\pi}
|\mathcal{T}_{H\rightarrow\gamma\gamma}|^{2}%
\end{equation}
and the decay coupling constant is:
\begin{equation}
g_{H\rightarrow\gamma\gamma}=\frac{\mathcal{T}_{H\rightarrow\gamma\gamma}}{e^2}.%
\end{equation}
As we have shown in \cite{costaD}, there is a good overall agreement between our results for these  quantities, calculated in the vacuum, and the experimental values \cite {pdb}. 


\section{Discussion and conclusions}
 
We will discuss our results for $\pi^0 (\eta)\rightarrow \gamma\gamma$ observables at finite $T$ and we  will show that, in spite of a different behavior of quantities relevant for the calculation of those observables, as compared to the case of finite density and zero temperature \cite{costaD}, the conclusions are qualitatively similar. As already mentioned, the phase transition with density is first order  and the mesons are still bound states beyond the critical density, although with a weaker coupling to the quarks; the phase transition with temperature is a crossover one and the mesons dissociate into $\bar q q$ at  the Mott transition point, where the meson masses cross the threshold $2M_u$ (Fig. 1, upper panel) and the meson-quark coupling constants decrease sharply, going  to zero (Fig. 2).

\begin{figure}[t]
  \begin{center}
    \includegraphics[width=0.36\textwidth]{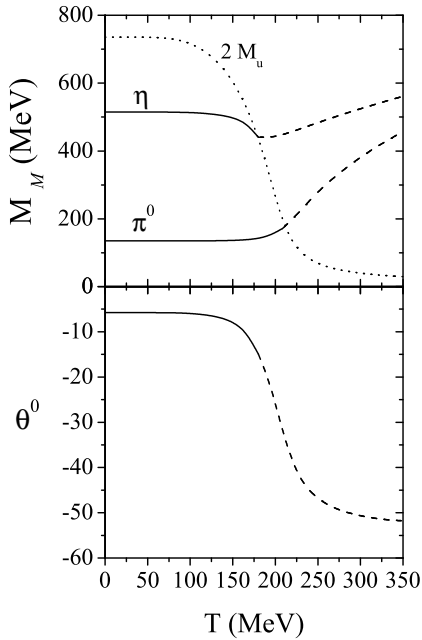}
  \end{center}
  \caption{Mesonic and quark masses (upper panel) 
  				and the mixing angle $\theta$ 
    			(lower panel) as functions of temperature.}
  \label{fig:mass}
\end{figure}
%
\begin{figure}[t]
	\begin{center}
    \hspace{-0.5cm}\includegraphics[width=0.50\textwidth]{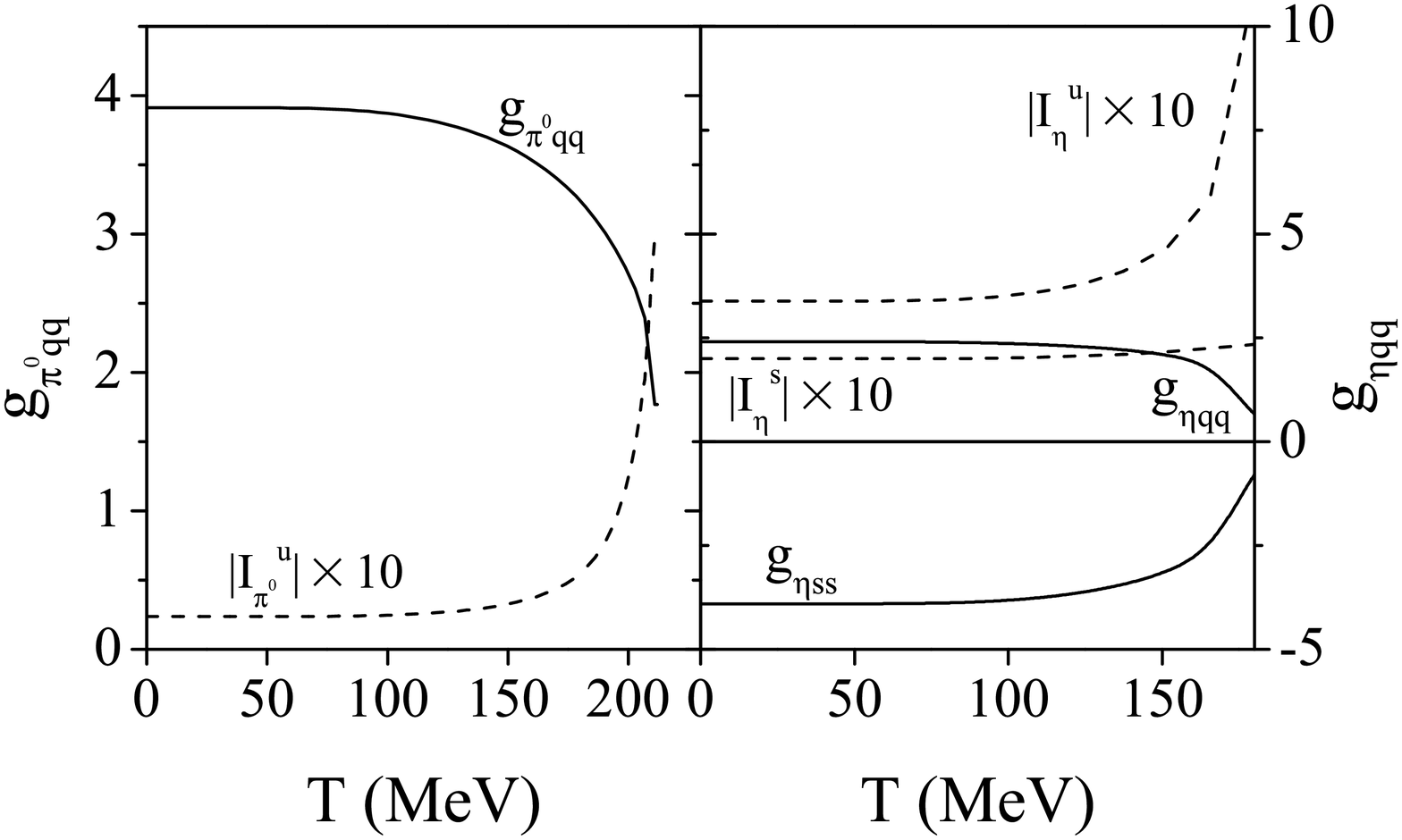}
	\end{center}
 	\caption{Meson--quark--quark coupling constants and integral $|{I^q}_H|\times 10$ as  									functions of temperature, for $\pi^0$ (left panel) and $\eta$ (right panel).}
 
 	\label{fig:acop}
\end{figure}
%
We plot ${\mathcal T}_{H\rightarrow \gamma\gamma}\,, \Gamma_{H\rightarrow \gamma\gamma}\,,g_{H\gamma\gamma}$, as functions of temperature for $\pi^0$  and for the $\eta$ (left panels of  Figs. 3 and 4, respectively). For the sake of comparison, we also plot these  quantities as functions of the baryonic density, in neutron matter, in the right panels of these figures.  Since we are mainly interested in discussing the behavior of these quantities up to the critical point, we plot our results only for $ T\leq T^{\pi^0}_{Mott}\approx 212$ MeV  and for $\rho_B\leq \rho_{cr}\approx 2.25 \rho_0$. 

The transition amplitude for the pion depends on the quark meson coupling constant, $g_{\pi^0 \bar u u}$ and on the integral $I^u_{\pi^0}(P_0=M_{\pi^0})$, that are plotted in Fig. 2, left panel. As discussed in Ref. \cite{tdavid,hashimoto,tklev}, at the Mott temperature there are threshold effects which cause a complicated behavior of the transition amplitude, a fact that does not occur at finite density where the behavior is smooth. While $g_{\pi^0 \bar u u}$ has a sharp decrease, $I^u_{\pi^0}$ has a sharp increase, the combination of these two effects leading to a smooth increase of the transition amplitude (see Fig. 3). The width has a sharp increase near the Mott temperature due to the increase of the pion mass, which is a manifestation of the tendency for the restoration of chiral symmetry. Let us remember that a sign for the restoration of this symmetry is that the mass of the pion increases with density and this meson becomes degenerated with $\sigma$ meson and the pion decay constant $f_\pi$ goes asymptotically to zero. 
\begin{figure}[t]
	\begin{center}
		\includegraphics[width=0.49\textwidth]{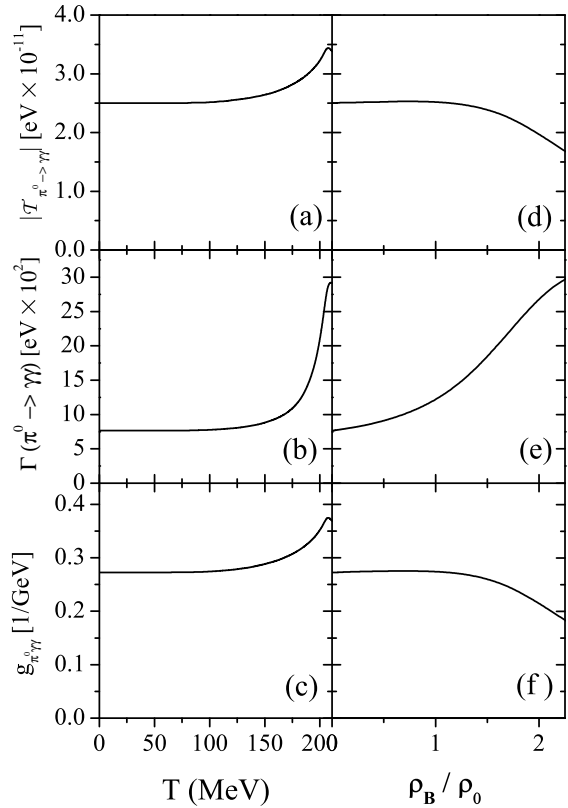}
	\end{center}
	\caption{The decay $\pi^0\rightarrow\gamma\gamma$: 
						transition amplitude, decay width and 
   					coupling constant as functions of temperature (a), b) 
   					and c)), and as functions of the density (d), e) and f)).}
	\label{fig:pi0}
\end{figure}

\begin{figure}[t]
	\begin{center}
		\includegraphics[width=0.47\textwidth]{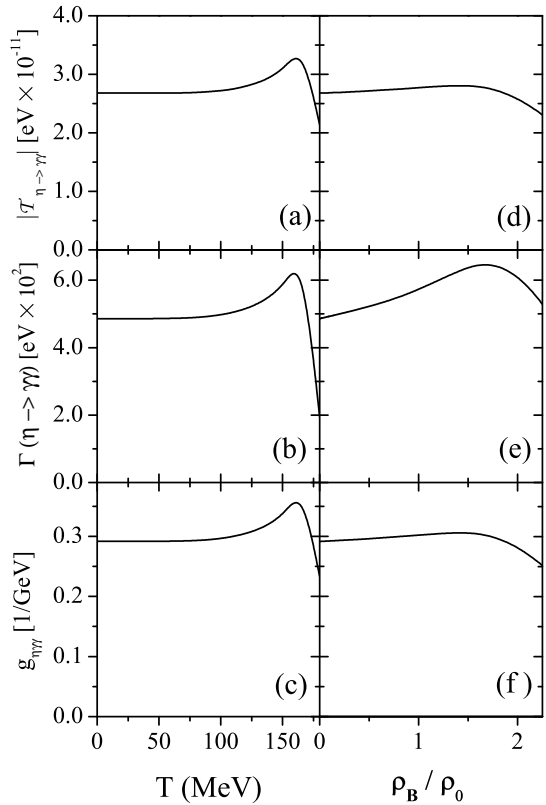}
	\end{center}
	\caption{The decay $\eta\rightarrow\gamma\gamma$: 
					transition amplitude, decay width and coupling 
					constant as functions of temperature (a), b) and 	c)), 
 					and as functions of the density (d), e) and f)).}
	\label{fig:eta}
\end{figure}

Concerning the $\eta \rightarrow \gamma\gamma$ decay (Fig. 4), although qualitatively similar to $\pi^0 \rightarrow \gamma\gamma$, it depends on the couplings to strange and non strange  quarks, $g_{\eta \bar s s}\,,g_{\eta \bar u u}$,  on the integrals $I^s_{\eta}(P_0=M_\eta),\,I^u_{\eta}(P_0=M_\eta)$ (Fig. 2, right panel) and also on the mixing angle $\theta (M_\eta)$, that we plot in Fig. 1, lower panel. The behavior of the mixing angle is very smooth, but the coupling constants and the integral referred to above show a behavior similar to the case of the pion. The main difference is that  the peak for the decay amplitude occurs before the Mott temperature for the $\eta$, $T^{\eta}_{Mott}\approx 180$ MeV.

In conclusion, we show that these anomalous decays are significantly affected by the medium, leading to an enhancement of the width around the Mott temperature. However,  this enhancement is probably not sufficient to lead to lifetimes shorter than the expected lifetime of the fireball.  Recent experimental results from PHENIX \cite{phenix} show that $\pi^0$ production is suppressed in the central region of Au$+$Au collisions as compared to the peripherical region. This means that $\pi^0 \rightarrow \gamma\gamma$ decay could only be interesting for experimental heavy-ion collisions at intermediate temperatures and densities. However, although the peak of the $\pi^0$ and $\eta$ widths are at  moderate temperature, the decays are probably observed only after freeze-out, since its life times, $\tau$, are still of the order of $10^{-17}$ s and $10^{-18}$ s, respectively, much larger than the expected lifetime of the fireball, $10^{-22}$ s. A similar conclusion was obtained for matter at finite density and zero temperature, although the behavior of the relevant observables is more smooth.  
 

\begin{center}
{\large Acknowledgments:}
\end{center}
Work supported by Grant No. SFRH/BD/3296/2000 (P. Costa), by Grant No. RFBR 03-01-00657, Centro de F\'{\i}sica Te\'orica and GTAE (Yu. Kalinovsky).



\end{document}